\documentstyle[preprint,aps]{revtex}
\tightenlines
\begin{document}
\draft
\title{A fractal-like structure for the fractional quantum Hall effect}

\author{Wellington da Cruz}
\address{Departamento de F\'{\i}sica,\\
 Universidade Estadual de Londrina, Caixa Postal 6001,\\
Cep 86051-970 Londrina, PR, Brazil\\
E-mail address: wdacruz@exatas.uel.br}
\date{\today}
\maketitle
\begin{abstract}

We have pursued in the literature a fractal-like structure 
for the fractional quantum Halll effect-FQHE which consider 
the Hausdorff dimension associated with the quantum mechanics paths 
and the spin of the particles or quasiparticles termed fractons. These objects 
carry rational or irrational values of spin and satisfy a fractal distribution 
function associated with a fractal von Neumann entropy. We show that our approach  
offers {\it a rationale} for all FQHE data 
including possible filling factors suggested by some authors. Our formulation is 
free of any empirical formula and this characteristic appears 
as a foundational insight for this FQHE-phenomenon. The connection between 
a geometrical parameter, the Hausdorff dimension $h$, associated 
with the quantum paths and the spin $s$ of particles, $h=2-2s$, 
$0 < s < \frac{1}{2}$, is 
a physical analogous to the fractal dimension formula, $\Delta(\Gamma)=2-H$, of the 
graph of the functions in the context of the fractal geometry, where $H$ 
is known as H\"older exponent, with $ 0 < H < 1$. We emphasize that our fractal 
approach to the fractional spin particles gives us a new perspective 
for charge-flux systems defined in two-dimensional multiply connected space.  
\end{abstract}

\pacs{PACS numbers: 71.10.Pm, 05.30.-d, 73.43.Cd, 05.30.Pr\\
Keywords:  Fractal distribution function; Fractal von Neumann entropy; 
Fractons; Fractional quantum Hall effect.}

\newpage

\section{Introduction}

The Fractal Geometry of Nature\cite{R1} is manifest in the context of the 
fractional quantum Hall effect\cite{R2} through a theoretical 
formulation introduced by us in the 
literature\cite{R3,R4,R5,R6,R7,R8,R9,R10}. Another viewpoint about 
the fractal approach to the FQHE was discussed in\cite{R11}. We have 
defined universal classes $h$ of 
particles or quasiparticles, termed fractons, which satisfy a 
fractal distribution function associated with a fractal von Neumann entropy. 
Such objects are charge-flux systems defined in two-dimensional 
multiply connected space and they carry rational or irrational values of spin. 
We have considered the Hausdorff dimension $h$, defined within the interval 
$1$$\;$$ < $$\;$$h$$\;$$ <$$\;$$ 2$ and associated with the fractal 
curves of the quantum mechanics particles. We have found an expression which relates 
the fractal dimension $h$ and the spin $s$ of the particles, 
$h=2-2s$, $0 < s < \frac{1}{2}$. This result has a mathematical 
analogous, in the branch of the fractal geometry, to the fractal dimension formula 
of the graph of the functions and given by: $\Delta(\Gamma)=2-H$, where $H$ is 
known as H\"older exponent, with $0 < H < 1$\cite{R12}. The fractal properties of the 
quantum paths can be extracted from the propagators of the particles 
in the momentum space\footnote{Another viewpoint about this discussion can 
be obtained via a gamma representation of the statistical weight given bellow\cite{R25}.}
\cite{R3,R13}. Our formula, when we consider 
the spin-statistics relation $\nu=2s$, is written as $h=2-\nu$, $0 < \nu < 1$. 
In this way, a fractal spectrum was defined taking into account a mirror symmetry:

\begin{eqnarray}
h-1&=&1-\nu,\;\;\;\; 0 < \nu < 1;\;\;\;\;\;\;\;\;h-1=
\nu-1,\;\;\;\;\;\;\; 1 <\nu < 2;\nonumber\\
h-1&=&3-\nu,\;\;\;\; 2 < \nu < 3;\;\;\;\;\;\;\;\;
 h-1=\nu-3,\;
\;\;\;\;\;\; 3 <\nu < 4;\;etc.
\end{eqnarray}

Now, given the statistical weight for these classes of fractons\cite{R3}

\begin{equation}
\label{e11}
{\cal W}[h,n]=\frac{\left[G+(nG-1)(h-1)\right]!}{[nG]!
\left[G+(nG-1)(h-1)-nG\right]!}
\end{equation}

and from the condition of the entropy be a maximum, we obtain 
the fractal distribution function

\begin{eqnarray}
\label{e.44} 
n[h]=\frac{1}{{\cal{Y}}[\xi]-h}.
\end{eqnarray}

The function ${\cal{Y}}[\xi]$ satisfies the equation 

\begin{eqnarray}
\label{e.4} 
\xi=\biggl\{{\cal{Y}}[\xi]-1\biggr\}^{h-1}
\biggl\{{\cal{Y}}[\xi]-2\biggr\}^{2-h},
\end{eqnarray}

\noindent with $\xi=\exp\left\{(\epsilon-\mu)/KT\right\}$. We understand the 
fractal distribution function as a quantum-geometrical 
description of the statistical laws of nature, 
since the quantum path is a fractal curve and this 
reflects the Heisenberg uncertainty principle. 

We can obtain for any class its distribution function considering the
Eqs.(\ref{e.44},\ref{e.4}). For example, 
the universal class $h=\frac{3}{2}$ with distinct values of spin 
$\biggl\{\frac{1}{4},\frac{3}{4},\frac{5}{4},\cdots\biggr\}_{h=\frac{3}{2}}$, 
has a specific fractal distribution

\begin{eqnarray}
n\left[\frac{3}{2}\right]=\frac{1}{\sqrt{\frac{1}{4}+\xi^2}}.
\end{eqnarray}

\noindent This result coincides with another 
one of the literature of fractional spin particles for 
the statistical parameter $\nu=\frac{1}{2}$\cite{R14}, 
however our interpretation is completely distinct\footnote{In particular, 
the authors in\cite{R14} never considered the possibility 
of to define universal class of particles for charge-flux systems 
defined in two-dimensional multiply connected space. We have here, 
sets of particles ( fractons )  with rational or irrational values of spin, 
satisfying a specific fractal distribution function. This idea works 
in the same way that fermions constitute a universal class 
of particles obeying the Fermi-Dirac distribution function. We emphasize 
that, in this fractal approach to the fractional spin particles, fractons 
realize the spin-statistics connection. On the other hand, the Hausdorff 
dimension associated to the quantum paths of particles with any value of spin 
can be obtained promptly.}. We also have
 
\begin{eqnarray}
\xi^{-1}=\biggl\{\Theta[{\cal{Y}}]\biggr\}^{h-2}-
\biggl\{\Theta[{\cal{Y}}]\biggr\}^{h-1},
\end{eqnarray}

\noindent where

\begin{eqnarray}
\Theta[{\cal{Y}}]=
\frac{{\cal{Y}}[\xi]-2}{{\cal{Y}}[\xi]-1}
\end{eqnarray}

\noindent is the single-particle partition function. 
We verify that the classes $h$ satisfy a duality symmetry defined by 
${\tilde{h}}=3-h$. So, fermions and bosons come as dual particles. 
As a consequence, we extract a fractal 
supersymmetry which defines pairs of particles $\left(s,s+\frac{1}{2}\right)$. 
This way, the fractal distribution function appears as 
a natural generalization of the fermionic and bosonic 
distributions for particles with braiding properties. Therefore, 
our approach is a unified formulation 
in terms of the statistics which each universal class of 
particles satisfies, from a unique expression 
we can take out any distribution function. In some sense , we can say that 
fermions are fractons of the class $h=1$ and  
bosons are fractons of the class $h=2$.

The free energy for particles in a given quantum state is expressed as

\begin{eqnarray}
{\cal{F}}[h]=KT\ln\Theta[{\cal{Y}}].
\end{eqnarray}

\noindent Hence, we find the average occupation number

\begin{eqnarray}
\label{e.h} 
n[h]&=&\xi\frac{\partial}{\partial{\xi}}\ln\Theta[{\cal{Y}}].
\end{eqnarray}

\noindent The fractal von Neumann entropy per state in terms of the 
average occupation number is given as\cite{R3,R4} 

\begin{eqnarray}
\label{e5}
{\cal{S}}_{G}[h,n]&=& K\left[\left[1+(h-1)n\right]\ln\left\{\frac{1+(h-1)n}{n}\right\}
-\left[1+(h-2)n\right]\ln\left\{\frac{1+(h-2)n}{n}\right\}\right]
\end{eqnarray}

\noindent and it is associated with the fractal distribution function Eq.(\ref{e.44}).

Now, as we can check, each universal class $h$ of particles, 
within the interval of definition has its entropy defined 
by the Eq.(\ref{e5}). Thus, for fractons of the self-dual class
$\biggl\{\frac{1}{4},
\frac{3}{4},\frac{5}{4},\cdots\biggr\}_{h=\frac{3}{2}}$, we obtain
  
\begin{eqnarray}
{\cal{S}}_{G}\left[\frac{3}{2}\right]=K\left\{(2+n)\ln\sqrt{\frac{2+n}{2n}}
-(2-n)\ln\sqrt{\frac{2-n}{2n}}\right\}. 
\end{eqnarray}

\noindent We have also introduced the topological concept of fractal index, 
which is associated with each class. As we saw, $h$ is a geometrical parameter 
related to the quantum paths of the particles and so, we define\cite{R5} 

\begin{equation}
\label{e.1}
i_{f}[h]=\frac{6}{\pi^2}\int_{\infty(T=0)}^{1(T=\infty)}
\frac{d\xi}{\xi}\ln\left\{\Theta[\cal{Y}(\xi)]\right\}.
\end{equation}

\noindent For the interval of the definition $ 1$$\;$$ \leq $$\;$$h$$\;$$ \leq $$\;$$ 2$, there 
exists the correspondence $0.5$$\;$$ 
\leq $$\;$$i_{f}[h]$$\;$$ \leq $$\;$$ 1$, which signalizes 
a connection between fractons and quasiparticles of the conformal field theories, 
in accordance with the unitary $c$$\;$$ <$$\;$$ 1$ 
representations of the central charge. For $\nu$ even it is defined by 

\begin{eqnarray}
\label{e.11}
c[\nu]=i_{f}[h,\nu]-i_{f}\left[h,\frac{1}{\nu}\right]
\end{eqnarray}

\noindent and for $\nu$ odd it is defined by 

\begin{eqnarray}
\label{e.12}
c[\nu]=2\times i_{f}[h,\nu]-i_{f}\left[h,\frac{1}{\nu}\right],
\end{eqnarray}

\noindent where $i_{f}[h,\nu]$ means the fractal 
index of the universal class $h$ which contains the particles 
with distinct values of spin. 
In another way, the central charge $c[\nu]$ can be obtained using the 
Rogers dilogarithm function, i.e. 

\begin{equation}
\label{e.16}
c[\nu]=\frac{L[x^{\nu}]}{L[1]},
\end{equation}

\noindent with $x^{\nu}=1-x$,$\;$ $\nu=0,1,2,3,etc.$ and 

\begin{equation}
L[x]=-\frac{1}{2}\int_{0}^{x}\left\{\frac{\ln(1-y)}{y}
+\frac{\ln y}{1-y}\right\}dy,\; 0 < x < 1.
\end{equation}

\noindent Thus, we have established a connection between fractal geometry and 
number theory, given that the dilogarithm function appears 
in this context, besides another branches of mathematics\cite{R15}. 

Such ideas can be applied in the context of the FQHE. This phenomenon is 
characterized  by the filling factor parameter $f$, and for 
each value of $f$ we have the 
quantization of Hall resistance and a superconducting state 
along the longitudinal direction of a planar system of electrons, which are
manifested by semiconductor doped materials, i.e., heterojunctions 
under intense perpendicular magnetic fields and lower 
temperatures\cite{R2}. 

The parameter $f$ is defined by $f=N\frac{\phi_{0}}{\phi}$, where 
$N$ is the electron number, 
$\phi_{0}$ is the quantum unit of flux and
$\phi$ is the flux of the external magnetic field throughout the sample. 
The spin-statistics relation is given by 
$\nu=2s=2\frac{\phi\prime}{\phi_{0}}$, where 
$\phi\prime$  is the flux associated with the charge-flux 
system which defines the fracton $(h,\nu)$. According to our approach 
there is a correspondence between $f$ and $\nu$, numerically $f=\nu$. 
This way, we verify that the filling factors 
experimentally observed  appear into the classes $h$ and from the definition of duality 
between the equivalence classes, we note that the FQHE occurs in pairs 
 of these dual topological quantum numbers.

\section{Fractal structure for the FQHE data}

In\cite{R10} we have considered recent experimental results 
and checked the validity of our approach. On the one hand, 
some papers have suggested other  
values for the filling factors. In this Letter, 
we show that these possible values and their duals can be obtained 
by our formulation which is free of any empirical formula, i.e., 
our approach has a foundational basis connecting a fractal 
parameter $h$ associated with the quantum paths and the spin $s$ of fractons. 
Besides this, these particles obey a fractal distribution 
function associated with a fractal von Neumann entropy, 
as we discussed earlier. On the other hand, 
the suggestion in\cite{R16} about {\it the fractal-like 
structure to a deeper understanding of FQHE} echoed just with our ideas 
advanced in the literature\cite{R3,R4,R5,R6,R7,R8,R9,R10}. 


Let us consider the filling factors suggested in\cite{R16,R17,R18}  
and the FQHE data\cite{R11,R19,R20,R21,R22,R23} together. Taking into account 
the fractal spectrum and the duality symmetry, we can write the sequence:

\begin{eqnarray}
&&{\bf \frac{4}{1}}\;\rightarrow\frac{19}{5}\rightarrow\;
\frac{34}{9}\rightarrow\frac{15}{4}\rightarrow\frac{26}{7}\rightarrow
\frac{11}{3}\rightarrow\;
\frac{18}{5}\rightarrow\frac{25}{7}\rightarrow
\frac{7}{2}\rightarrow\frac{24}{7}\rightarrow
\frac{17}{5}\rightarrow\frac{10}{3}\rightarrow\;\;
\frac{23}{7}\rightarrow\nonumber\\
&&\frac{13}{4}\rightarrow
\frac{29}{9}\rightarrow\frac{16}{5}\rightarrow\;
{\bf \frac{3}{1}}\rightarrow\;\frac{14}{5}\rightarrow
\frac{25}{9}\rightarrow\;\frac{11}{4}\rightarrow
\frac{19}{7}\rightarrow\frac{8}{3}\rightarrow
\frac{13}{5}\rightarrow\frac{18}{7}\rightarrow\;
\frac{23}{9}\rightarrow\;\frac{28}{11}
\rightarrow\nonumber\\
&&\frac{33}{13}\rightarrow\;
\frac{5}{2}\rightarrow\;\frac{32}{13}\rightarrow
\frac{27}{11}\rightarrow\frac{22}{9}\rightarrow
\frac{17}{7}\rightarrow\;\frac{12}{5}\rightarrow\;
\frac{7}{3}\rightarrow\frac{16}{7}\rightarrow\;\frac{9}{4}
\rightarrow\;\frac{20}{9}\rightarrow
\frac{11}{5}
\rightarrow\;\;{\bf \frac{2}{1}}\rightarrow\nonumber\\
&&\frac{15}{8}
\rightarrow\frac{43}{23}\rightarrow\frac{28}{15}\rightarrow
\frac{13}{7}\rightarrow\;\frac{11}{6}\rightarrow
\frac{9}{5}\rightarrow\;\;\frac{16}{9}\rightarrow\;
\frac{7}{4}\rightarrow\frac{19}{11}\rightarrow
\frac{12}{7}\rightarrow\frac{17}{10}\rightarrow\;
\frac{5}{3}\rightarrow\;\;\;\frac{8}{5}\rightarrow\nonumber\\
&&\frac{11}{7}\rightarrow\frac{14}{9}\rightarrow
\frac{17}{11}\rightarrow\frac{20}{13}\rightarrow\;
\frac{3}{2}\rightarrow\;\frac{19}{13}\rightarrow\;
\frac{16}{11}\rightarrow\frac{13}{9}\rightarrow
\frac{10}{7}\rightarrow\frac{7}{5}\rightarrow\;\;
\frac{4}{3}\rightarrow\;\frac{13}{10}\rightarrow\;\;
\frac{9}{7}\rightarrow\nonumber\\
&&\frac{14}{11}\rightarrow\;
\frac{5}{4}\rightarrow\;\frac{11}{9}\rightarrow\;
\frac{6}{5}\rightarrow\;\;\frac{7}{6}\rightarrow\;
\frac{8}{7}\rightarrow\;\;\frac{17}{15}\rightarrow
\frac{26}{23}\rightarrow\;\frac{9}{8}\rightarrow\;
{\bf \frac{1}{1}}\;\;\rightarrow\frac{8}{9}\rightarrow\;\;
\frac{15}{17}\rightarrow\;
\frac{7}{8}\rightarrow\nonumber\\
&&\frac{13}{15}\rightarrow\;
\frac{6}{7}\rightarrow\;\frac{11}{13}\rightarrow
\frac{16}{19}\rightarrow\;\frac{5}{6}\rightarrow
\frac{19}{23}\rightarrow\;\;\frac{14}{17}\rightarrow
\frac{9}{11}\rightarrow\frac{22}{27}\rightarrow
\frac{13}{16}\rightarrow\frac{17}{21}\rightarrow
\frac{21}{26}\rightarrow\;
\frac{25}{31}\rightarrow\nonumber\\
&&\frac{4}{5}\rightarrow\;\frac{23}{29}\rightarrow\;
\frac{19}{24}\rightarrow\frac{15}{19}\rightarrow
\frac{11}{14}\rightarrow\frac{18}{23}\rightarrow\;\;
\frac{7}{9}\rightarrow\;\frac{10}{13}\rightarrow
\frac{13}{17}\rightarrow\frac{16}{21}\rightarrow
\frac{19}{25}\rightarrow\;\frac{3}{4}\rightarrow\;
\frac{17}{23}\rightarrow\nonumber\\
&&\frac{14}{19}\rightarrow\frac{11}{15}\rightarrow
\frac{8}{11}\rightarrow\;\frac{5}{7}\rightarrow\;
\frac{12}{17}\rightarrow\frac{7}{10}\rightarrow\;
\frac{16}{23}\rightarrow\frac{9}{13}\rightarrow
\frac{20}{29}\rightarrow\frac{11}{16}\rightarrow
\frac{24}{35}\rightarrow\frac{13}{19}\rightarrow\;
\frac{2}{3}\rightarrow\\
&&\frac{11}{17}\rightarrow
\frac{20}{31}\rightarrow\frac{9}{14}\rightarrow
\frac{16}{25}\rightarrow\frac{7}{11}\rightarrow
\frac{12}{19}\rightarrow\;\frac{17}{27}\rightarrow\;
\frac{5}{8}\rightarrow\;\frac{18}{29}\rightarrow\;
\frac{13}{21}\rightarrow\frac{8}{13}\rightarrow
\frac{3}{5}\;\rightarrow\frac{16}{27}\rightarrow\nonumber\\
&&\frac{29}{49}\rightarrow
\frac{13}{22}\rightarrow\frac{23}{39}\rightarrow
\frac{10}{17}\rightarrow\frac{17}{29}\rightarrow
\frac{24}{41}\rightarrow\;\frac{7}{12}\rightarrow
\frac{25}{43}\rightarrow\frac{18}{31}\rightarrow\;
\frac{11}{19}\rightarrow\;\frac{4}{7}\rightarrow
\frac{21}{37}\rightarrow\frac{38}{67}\rightarrow\nonumber\\
&&\frac{17}{30}\rightarrow\frac{30}{53}\rightarrow
\frac{13}{23}\rightarrow\frac{22}{39}\rightarrow
\frac{31}{55}\rightarrow\frac{9}{16}\rightarrow\;
\frac{32}{57}\rightarrow\frac{23}{41}\rightarrow
\frac{14}{25}\rightarrow\;\frac{5}{9}\rightarrow
\;\frac{6}{11}\rightarrow\frac{7}{13}\rightarrow
\frac{8}{15}\rightarrow\nonumber\\
&&\frac{9}{17}\rightarrow\frac{10}{19}\rightarrow
\frac{11}{21}\rightarrow\frac{1}{2}\rightarrow\;\;
\frac{10}{21}\rightarrow\frac{9}{19}\rightarrow
\frac{8}{17}\rightarrow\frac{7}{15}\rightarrow
\frac{6}{13}\rightarrow\;\frac{5}{11}\rightarrow\;
\frac{4}{9}\rightarrow\;\frac{11}{25}\rightarrow
\frac{18}{41}\rightarrow\nonumber\\
&&\frac{25}{57}\rightarrow\frac{7}{16}\rightarrow
\frac{24}{55}\rightarrow\frac{17}{39}\rightarrow
\frac{10}{23}\rightarrow\;\frac{23}{53}\rightarrow
\frac{13}{30}\rightarrow\frac{29}{67}\rightarrow
\frac{16}{37}\rightarrow\;\frac{3}{7}\rightarrow\;
\frac{8}{19}\rightarrow\frac{13}{31}\rightarrow
\frac{18}{43}\rightarrow\nonumber\\
&&\frac{5}{12}\rightarrow\frac{17}{41}\rightarrow
\frac{12}{29}\rightarrow\frac{7}{17}\rightarrow
\frac{16}{39}\rightarrow\;\frac{9}{22}\rightarrow
\frac{20}{49}\rightarrow\frac{11}{27}\rightarrow\;
\frac{2}{5}\rightarrow\;\frac{5}{13}\rightarrow
\frac{8}{21}\rightarrow\frac{11}{29}\rightarrow\;
\frac{3}{8}\rightarrow\nonumber\\
&&\frac{10}{27}\rightarrow\frac{7}{19}\rightarrow
\frac{4}{11}\rightarrow\frac{9}{25}\rightarrow
\frac{5}{14}\rightarrow\;\frac{11}{31}\rightarrow
\frac{6}{17}\rightarrow\;\frac{1}{3}\rightarrow\;
\frac{6}{19}\rightarrow\frac{11}{35}\rightarrow
\frac{5}{16}\rightarrow\frac{9}{29}\rightarrow
\frac{4}{13}\rightarrow\nonumber\\
&&\frac{7}{23}\rightarrow\frac{3}{10}\rightarrow
\frac{5}{17}\rightarrow\frac{2}{7}\rightarrow\;\;
\frac{3}{11}\rightarrow\;\frac{4}{15}\rightarrow
\frac{5}{19}\rightarrow\frac{6}{23}\rightarrow\;
\frac{1}{4}\rightarrow\;\frac{6}{25}\rightarrow
\frac{5}{21}\rightarrow\frac{4}{17}\rightarrow
\frac{3}{13}\rightarrow\nonumber\\
&&\frac{2}{9}\rightarrow\;\;\frac{5}{23}\rightarrow
\frac{3}{14}\rightarrow\frac{4}{19}\rightarrow
\frac{5}{24}\rightarrow\frac{6}{29}\rightarrow\;
\frac{1}{5}\rightarrow\;\frac{6}{31}\rightarrow\;
\frac{5}{26}\rightarrow\frac{4}{21}\rightarrow
\frac{3}{16}\rightarrow\frac{5}{27}\rightarrow
\frac{2}{11}\rightarrow\nonumber\\
&&\frac{3}{17}\rightarrow
\frac{4}{23}\rightarrow\;\frac{1}{6}\rightarrow
\frac{3}{19}\rightarrow\frac{2}{13}\rightarrow\;\;
\frac{1}{7}\rightarrow\;\frac{2}{15}\rightarrow\;
\frac{1}{8}\rightarrow\;\frac{2}{17}\rightarrow\;\;\frac{1}{9}.\nonumber 
\end{eqnarray}


\noindent Thus, we identify dual pairs of 
filling factors observed

\begin{eqnarray} 
\label{e.22}
(\nu,\tilde{\nu})&=&\left(\frac{1}{3},\frac{2}{3}\right), 
\left(\frac{5}{3},\frac{4}{3}\right), \left(\frac{1}{5},\frac{4}{5}\right), 
\left(\frac{2}{7},\frac{5}{7}\right),\left(\frac{2}{9},\frac{7}{9}\right), 
\left(\frac{2}{5},\frac{3}{5}\right),\left(\frac{7}{3},\frac{8}{3}\right),\nonumber\\
&&\left(\frac{3}{7},\frac{4}{7}\right), 
\left(\frac{4}{9},\frac{5}{9}\right), \left(\frac{8}{5},\frac{7}{5}\right),
\left(\frac{6}{13},\frac{7}{13}\right),\left(\frac{5}{11},\frac{6}{11}\right),
\left(\frac{10}{7},\frac{11}{7}\right),\\
&&\left(\frac{4}{11},\frac{7}{11}\right),
\left(\frac{13}{4},\frac{15}{4}\right), \left(\frac{14}{5},\frac{11}{5}\right),
\left(\frac{9}{4},\frac{11}{4}\right),\left(\frac{19}{7},\frac{16}{7}\right) 
etc.\nonumber
\end{eqnarray}

\noindent and other ones to be verified as 

\begin{eqnarray} 
(\nu,{\tilde{\nu}})&=&\left(\frac{2}{11},\frac{9}{11}\right),
\left(\frac{3}{19},\frac{16}{19}\right), 
\left(\frac{3}{17},\frac{14}{17}\right), \left(\frac{2}{13},\frac{11}{13}\right), 
\left(\frac{1}{7},\frac{6}{7}\right), \left(\frac{2}{15},\frac{13}{15}\right),\\
&&\left(\frac{5}{13},\frac{8}{13}\right),
\left(\frac{4}{13},\frac{9}{13}\right),\left(\frac{5}{17},\frac{12}{17}\right), 
\left(\frac{26}{7},\frac{23}{7}\right),\left(\frac{19}{5},\frac{16}{5}\right), 
\left(\frac{34}{9},\frac{29}{9}\right) etc.\nonumber
\end{eqnarray}

Here, we observe that our approach, in terms of equivalence 
classes for the filling factors, embodies 
the structure of the modular group as discussed in the literature
\cite{R3,R24} and the quantum Hall transitions satisfy some 
properties related with the Farey sequences of rational numbers. The 
transitions allowed are those generated by the condition
 $\mid p_{2}q_{1}
-p_{1}q_{2}\mid=1$, 
with $\nu_{1}=\frac{p_{1}}{q_{1}}$ and $\nu_{2}=
\frac{p_{2}}{q_{2}}$.  Thus, we verify distinct possible transitions for the 
fractional quantum Hall effect. We define families of 
universality classes to them, for example, consider the group I and II:\\
\\

{\bf Group I}

\begin{eqnarray}
&&\biggl\{\frac{1}{5},\frac{9}{5},\frac{11}{5},\frac{19}{5},
\cdots\biggr\}_{h=\frac{9}{5}}\rightarrow\;\;
\biggl\{\frac{2}{9},\frac{16}{9},\frac{20}{9},\frac{34}{9},
\cdots\biggr\}_{h=\frac{16}{9}}\rightarrow
\biggl\{\frac{1}{4},\frac{7}{4},\frac{9}{4},\frac{15}{4},
\cdots\biggr\}_{h=\frac{7}{4}}\rightarrow\nonumber\\ 
&&\biggl\{\frac{2}{7},\frac{12}{7},\frac{16}{7},\frac{26}{7},
\cdots\biggr\}_{h=\frac{12}{7}}\rightarrow 
\biggl\{\frac{1}{3},\frac{5}{3},\frac{7}{3},\frac{11}{3},
\cdots\biggr\}_{h=\frac{5}{3}}\rightarrow\;\;\;\;
\biggl\{\frac{2}{5},\frac{8}{5},\frac{12}{5},\frac{18}{5},
\cdots\biggr\}_{h=\frac{8}{5}}\rightarrow\nonumber\\
&&\biggl\{\frac{3}{7},\frac{11}{7},\frac{17}{7},\frac{25}{7},
\cdots\biggr\}_{h=\frac{11}{7}}
\rightarrow\biggl\{\frac{1}{2},\frac{3}{2},
\frac{5}{2},\frac{7}{2},\cdots\biggr\}_{h=\frac{3}{2}}
\rightarrow\biggl\{\frac{4}{7},\frac{10}{7},\frac{18}{7},\frac{24}{7},
\cdots\biggr\}_{h=\frac{10}{7}}\rightarrow\\
&&\biggl\{\frac{3}{5},\frac{7}{5},\frac{13}{5},\frac{17}{5},
\cdots\biggr\}_{h=\frac{7}{5}}\rightarrow\;\;\;
\biggl\{\frac{2}{3},\frac{4}{3},\frac{8}{3},\frac{10}{3},
\cdots\biggr\}_{h=\frac{4}{3}}
\rightarrow\biggl\{\frac{5}{7},\frac{9}{7},\frac{21}{7},\frac{23}{7},
\cdots\biggr\}_{h=\frac{9}{7}}
\rightarrow\nonumber\\
&&\biggl\{\frac{3}{4},\frac{5}{4},\frac{11}{4},\frac{13}{4},
\cdots\biggr\}_{h=\frac{5}{4}}\rightarrow\;\;\;
\biggl\{\frac{7}{9},\frac{11}{9},\frac{25}{9},
\frac{29}{9},\cdots\biggr\}_{h=\frac{11}{9}}
\rightarrow
\biggl\{\frac{4}{5},\frac{6}{5},\frac{14}{5},\frac{16}{5},
\cdots\biggr\}_{h=\frac{6}{5}}.\nonumber
\end{eqnarray}


\noindent {\bf Group II}

\begin{eqnarray}
&&\biggl\{\frac{4}{5},\frac{6}{5},\frac{14}{5},\frac{16}{5},
\cdots\biggr\}_{h=\frac{6}{5}}\rightarrow\;\;\;\;
\biggl\{\frac{7}{9},\frac{11}{9},\frac{25}{9},\frac{29}{9},
\cdots\biggr\}_{h=\frac{11}{9}}\rightarrow\;\;
 \biggl\{\frac{3}{4},\frac{5}{4},\frac{11}{4},\frac{13}{4},
\cdots\biggr\}_{h=\frac{5}{4}}\rightarrow\nonumber\\
&&\biggl\{\frac{5}{7},\frac{9}{7},\frac{19}{7},\frac{23}{7},
\cdots\biggr\}_{h=\frac{9}{7}}
\rightarrow\;\;\;\;
\biggl\{\frac{2}{3},\frac{4}{3},\frac{8}{3},\frac{10}{3},
\cdots\biggr\}_{h=\frac{4}{3}}
\rightarrow\;\;\;\;\;\;\; 
\biggl\{\frac{3}{5},\frac{7}{5},\frac{13}{5},\frac{17}{5},
\cdots\biggr\}_{h=\frac{7}{5}}
\rightarrow\nonumber\\
&&\biggl\{\frac{4}{7},\frac{10}{7},\frac{18}{7},\frac{24}{7},
\cdots\biggr\}_{h=\frac{10}{7}}
\rightarrow\;\;
\biggl\{\frac{5}{9},\frac{13}{9},\frac{23}{9},\frac{31}{9},
\cdots\biggr\}_{h=\frac{13}{9}}\rightarrow\;  
\biggl\{\frac{6}{11},\frac{16}{11},\frac{28}{11},\frac{38}{11},
\cdots\biggr\}_{h=\frac{16}{11}}\rightarrow\nonumber\\
&&\biggl\{\frac{7}{13},\frac{19}{13},\frac{33}{13},\frac{45}{13},
\cdots\biggr\}_{h=\frac{19}{13}}
\rightarrow
\biggl\{\frac{1}{2},\frac{3}{2},\frac{5}{2},\frac{7}{2},
\cdots\biggr\}_{h=\frac{3}{2}}
\rightarrow\;\;\;\;\;\;\;\; 
\biggl\{\frac{6}{13},\frac{20}{13},\frac{32}{13},\frac{46}{13},
\cdots\biggr\}_{h=\frac{20}{13}}
\rightarrow\\
&&\biggl\{\frac{5}{11},\frac{17}{11},\frac{27}{11},\frac{39}{11},
\cdots\biggr\}_{h=\frac{17}{11}}
\rightarrow
\biggl\{\frac{4}{9},\frac{14}{9},\frac{22}{9},\frac{32}{9},
\cdots\biggr\}_{h=\frac{14}{9}}\rightarrow\;\;  
\biggl\{\frac{3}{7},\frac{11}{7},\frac{17}{7},\frac{25}{7},
\cdots\biggr\}_{h=\frac{11}{7}}\rightarrow\nonumber\\
&&\biggl\{\frac{2}{5},\frac{8}{5},\frac{12}{5},\frac{18}{5},
\cdots\biggr\}_{h=\frac{8}{5}}
\rightarrow\;\;\;\;\;
\biggl\{\frac{1}{3},\frac{5}{3},\frac{7}{3},\frac{11}{3},
\cdots\biggr\}_{h=\frac{5}{3}}
\rightarrow\;\;\;\;\;\; 
\biggl\{\frac{2}{7},\frac{12}{7},\frac{16}{7},\frac{26}{7},
\cdots\biggr\}_{h=\frac{12}{7}}
\rightarrow\nonumber\\
&&\biggl\{\frac{1}{4},\frac{7}{4},\frac{9}{4},\frac{15}{4},
\cdots\biggr\}_{h=\frac{7}{4}}\rightarrow\;\;\;\;\;\;
\biggl\{\frac{2}{9},\frac{16}{9},\frac{20}{9},\frac{34}{9},
\cdots\biggr\}_{h=\frac{16}{9}}\rightarrow\;\;
\biggl\{\frac{1}{5},\frac{9}{5},\frac{11}{5},\frac{19}{5},
\cdots\biggr\}_{h=\frac{9}{5}}
.\nonumber
\end{eqnarray}

\section{Conclusions}

We have verified that our approach to the FQHE 
reproduces all experimental data and can predicting 
the occurrence of this phenomenon 
for other filling factors. According to our formulation, the topological 
character of these quantum numbers is related with the Hausdorff 
dimension of the quantum paths of fractons. For that, we have obtained 
a physical analogous to the mathematical one formula 
of the Hausdorff dimension associated with fractal curves. The FQHE occurs in pairs 
of dual topological quantum numbers filling factors. The 
foundational basis of our theoretical formulation is 
free of any empirical formula and this characteristic 
constitutes the great difference between our insight 
and others of the literature. Besides this, we have obtained other results, such as
\cite{R3,R4,R5,R6,R7,R8,R9,R10}: a relation between the 
fractal parameter and the Rogers dilogarithm function, through the 
concept of fractal index, which is defined 
in terms of the partition function associated with each 
universal class of particles; a connection between the fractal 
parameter $h$ and the Farey 
sequences of rational numbers. Farey series $F_{n}$ of order 
$n$ is the increasing sequence of 
irreducible fractions in the range $0-1$ whose 
denominators do not exceed $n$. We have the following 

{\bf Theorem}\cite{R8}: {\it The elements of the Farey series $F_{n}$ 
of the order $n$, belong to the fractal sets, whose Hausdorff 
dimensions are the second fractions of the fractal sets. The 
Hausdorff dimension has values within the interval 
$1$$\;$$ < $$\;$$h$$\;$$ <$$\;$$ 2$ and these ones 
are associated with fractal curves.}

Along this discussion we have established a connection between the FQHE 
and some concepts of the fractal geometry. We believe that our formulation 
sheds some light on that phenomenon. In another direction, a quantum 
computing in terms of fractons can be also defined and in this way, via
the concept of entanglement fractal von Neumann entropy, we have observed that 
different Hall states have equivalent entanglement content\cite{R25} . This is so 
because these Hall states belong to the same universality class 
of the quantum Hall transitions which are labelled just by the Hausdorff dimension. 
Also, we noted that some properties of the FQHE are independent of 
dynamical aspects or others details of the system: 
certain peculiarities of the FQHE can be extracted considering 
global properties as the modular group and the fractal dimension, 
which are implicit in all our theoretical formulation. Finally, 
the comments in\cite{R16} about a possible {\it fractal-like 
structure to a deeper understanding of the FQHE}, maybe can start with our ideas.

\end{document}